\documentclass[12pt]{article}

\usepackage{amsfonts,amssymb,chet,dsfont,mathrsfs,pgfplots,tikz,graphicx}

\newcommand{\Mbol}{M_\text{bol}}
\newcommand{\Lsun}{L_{\odot}}
\newcommand{\Rsun}{R_{\odot}}

\usetikzlibrary{decorations.markings,arrows,calc}
\tikzset{>=latex,->/.style={decoration={markings,mark=at position 1 with {\arrow[scale=1.5]{>}}},postaction={decorate}}}
\tikzset{->-/.style={decoration={markings,mark=at position 0.6 with {\arrow[scale=1.5]{>}}},postaction={decorate}}}
\tikzset{-<-/.style={decoration={markings,mark=at position 0.6 with {\arrow[scale=1.5]{<}}},postaction={decorate}}}
\preprint{BONN-TH-2013-04\\
CERN-PH-TH/2013-039\\
SU-ITP-13/02\vspace{-1.6cm}
}

\title{White Dwarfs constrain Dark Forces}

\author{Herbert K. Dreiner$^{\ast,}$\email{dreiner@th.physik.uni-bonn.de},
Jean-Fran\c{c}ois Fortin$^{\dagger,\$,}$\email{jean-francois.fortin@cern.ch},
Jordi Isern$^{\S,\ddag,}$\email{isern@aliga.ieec.uab.es} and
Lorenzo Ubaldi$^{\ast,}$\email{ubaldi@th.physik.uni-bonn.de}
}

\affiliation{$^\ast$Physikalisches Institut der Universit\"{a}t Bonn, Nussallee 12, D-53115 Bonn, Germany\\
$^\dagger$Theory Division, Department of Physics, CERN, CH-1211 Geneva 23, Switzerland\\
$^\$$Stanford Institute for Theoretical Physics, Department of Physics, Stanford University, Stanford, CA 94305, USA\\
$^\S$ Institut de Ci\`encies de l'Espai ICE(CSIC/IEEC), Campus UAB, 08193 Bellaterra, Spain\\
$^\ddag$ Institut d'Estudis Espacials de Catalunya (IEEC)
}

\abstract{The white dwarf luminosity function, which provides information about their cooling, has been measured with high
precision in the past few years.  Simulations that include well known Standard Model physics give a good fit to the data.  This leaves little room 
for new physics and makes these astrophysical objects a good laboratory for testing models beyond the Standard Model.  It has already been 
suggested that white dwarfs might provide some evidence for the existence of axions.  In this work we study the constraints that the white dwarf 
luminosity function puts on physics beyond the Standard Model involving new light particles (fermions or bosons) that can be pair-produced in 
a white dwarf and then escape to contribute to its cooling.  We show, in particular, that we can severely constrain the parameter space of models
with dark forces and light hidden sectors (lighter than a few tens of keV).  The bounds we find are often more competitive than those from current lab searches 
and those expected from most future searches.
}

\date{March 2013} %Uncomment this line for month to be fixed

\begin{document}

\maketitle

%%%%%%%%%%%%%%%%%%%%%%%%%%%%%%%%%%%%%%%%%%%%%%%%%%%%%%%%%%%%%%%%%%%%%%%
%%%%%%%%%%%%%%%%%%%%%%%%%%%%%%%%%%%%%%%%%%%%%%%%%%%%%%%%%%%%%%%%%%%%%%%
%%%%%%%%%%%%%%%%%%%%%%%%%%
%%%%%%%%%%%%%%%%%%%%%%%%%%%%%%%%%%%%%%%%%%%%%%%%%%%%%%%%%%%%%%%%%%%%%%%
%%%%%%%%%%%%%%%%%%%%%%%%%%%%%%%%%%%%%%%%%%%%%%%%%%%%%%%%%%%%%%%%%%%%%%%
%%%%%%%%%%%%%%%%%%%%%%%%%%

\section{Introduction}\label{Intro}

White dwarfs (WDs) are simple astrophysical objects whose cooling law is well understood.  This fact makes them a good laboratory for testing 
new models of particle physics.  Many such models predict the existence of light bosons or light fermions that interact very weakly with 
regular matter.  If these new particles are produced in a WD, they will typically escape and accelerate the cooling of the star.  Thus, determining 
the cooling law from astrophysical observations can be translated into constraints on particle physics beyond the Standard Model (BSM)~\cite{Raffelt:1996wa}.

We first give a brief review of WD cooling.  
Formally, the cooling evolution of white dwarfs can be written as:
\eqn{
L_\gamma+L_\nu+L_x=-\int_0^{M_\text{WD}}C_\text{v}\frac{dT}{dt}\,dm-\int_0^{M_\text{WD}}T\left(\frac{\partial P}{\partial T}\right)_{V,X_0}\frac{dV}{dt}\,dm+(l_\text{s}+\epsilon_\text{g})\frac{dM_\text{s}}{dt}+\dot\epsilon_x,
}[lwd]
where $L_\gamma$ and $L_\nu$ represent the photon and neutrino luminosities (energy per unit time). The first term on the r.h.s. is the well 
known contribution of the heat capacity of the star to the total luminosity, the second one represents the contribution of the
change of volume. It is in general small since only the thermal part
of the electronic pressure, the ideal part of the ions and the Coulomb
terms other than the Madelung term contribute~\cite{isern97}. The third term represents the contribution of
the latent heat and gravitational readjustement of the white dwarf to the total luminosity at freezing.  Finally, $L_x$ and $\dot\epsilon_x$ represent 
any extra energy sink or source of energy respectively. For many applications, this equation can be easily evaluated assuming an isothermal, 
almost completely degenerate core containing the bulk of the mass, surrounded by a thin, nondegenerate envelope.\footnote{The 
isothermal approximation is not valid when neutrinos are dominant, however the results are still reasonably good and provide a reasonable 
estimate of the luminosity. The ions do not follow the ideal gas law but the equation of state of a Coulomb plasma---for instance, in the region of 
interest the specific heat approaches the Dulong--Petit law---and crystallizes at low temperatures, around bolometric magnitude $12-13$, 
depending on the mass of the star.}

The evolution of white dwarfs can be tested through the luminosity function (LF), $n(l)$, which is defined as the  number of white dwarfs of a  
given luminosity or bolometric magnitude\footnote{The bolometric magnitude and the luminosity are related through
\eqn{
\Mbol = -2.5 \log_\text{10}(L/\Lsun)+4.74.
}[EqMbol]}
per unit of magnitude interval and unit volume:
\eqn{
n(l)\equiv\int^{M_\text{s}}_{M_\text{i}}\,\Phi(M)\,\Psi(\tau)\tau_\text{cool}(l,M)\,dM
}[ewdlf]
where
\eqn{
\tau\equiv T_G-t_\text{cool}(l,M)-t_\text{PS}(M).
}[bc]
$T_G$  is  the  age  of the Galaxy, $l\equiv-\log(L/\Lsun)$, $M$ is the  mass of the parent star (for convenience all white dwarfs
are labeled  with the mass of the main sequence  progenitor), $t_\text{cool}$ is the cooling time down to luminosity $l$, $\tau_\text{cool}=dt/d\Mbol$ is the characteristic cooling time, $M_\text{s}$ is the maximum mass of a main sequence star able to produce a white dwarf, 
and $M_\text{i}$ is the  minimum mass of the main sequence stars able to produce a white dwarf of luminosity $l$, and $t_\text{PS}$ is the lifetime of the progenitor of the white dwarf.  $\Phi(M)$ is the initial mass function, \textit{i.e.} the number of main sequence stars of mass $M$ that are born per unit mass, and $\Psi(t)$ is the star formation rate, \textit{i.e.} the mass per unit time and volume converted into stars.  So the product $\Phi(M)\Psi(\tau)$ is the number of main sequence stars that were born at the right moment to produce a white dwarf of luminosiy $l$ now.  Since the total density of white dwarfs is not well known, the computed
luminosity function is usually normalized to the bin with the smallest error bar, traditionally the one with $l=3$, in order to compare theory with observations.

The star formation rate is not known, but fortunately the bright part of Eq.~\ewdlf satisfies~\cite{Isern:2008fs}:
\eqn{
n(l)\propto\langle\tau_\text{cool}\rangle\int{\,\Phi(M)\Psi(\tau)\,dM}.
}
If $\Psi$ is a well behaved function and $T_G$ is large enough, the lower limit of the integral is not sensitive to the luminosity, and its value is 
absorbed by the normalization procedure in such a way that the shape of the luminosity function only depends on the averaged characteristic 
cooling time of white dwarfs.
\begin{figure}[!t]
\centering
\includegraphics[width=0.7\textwidth]{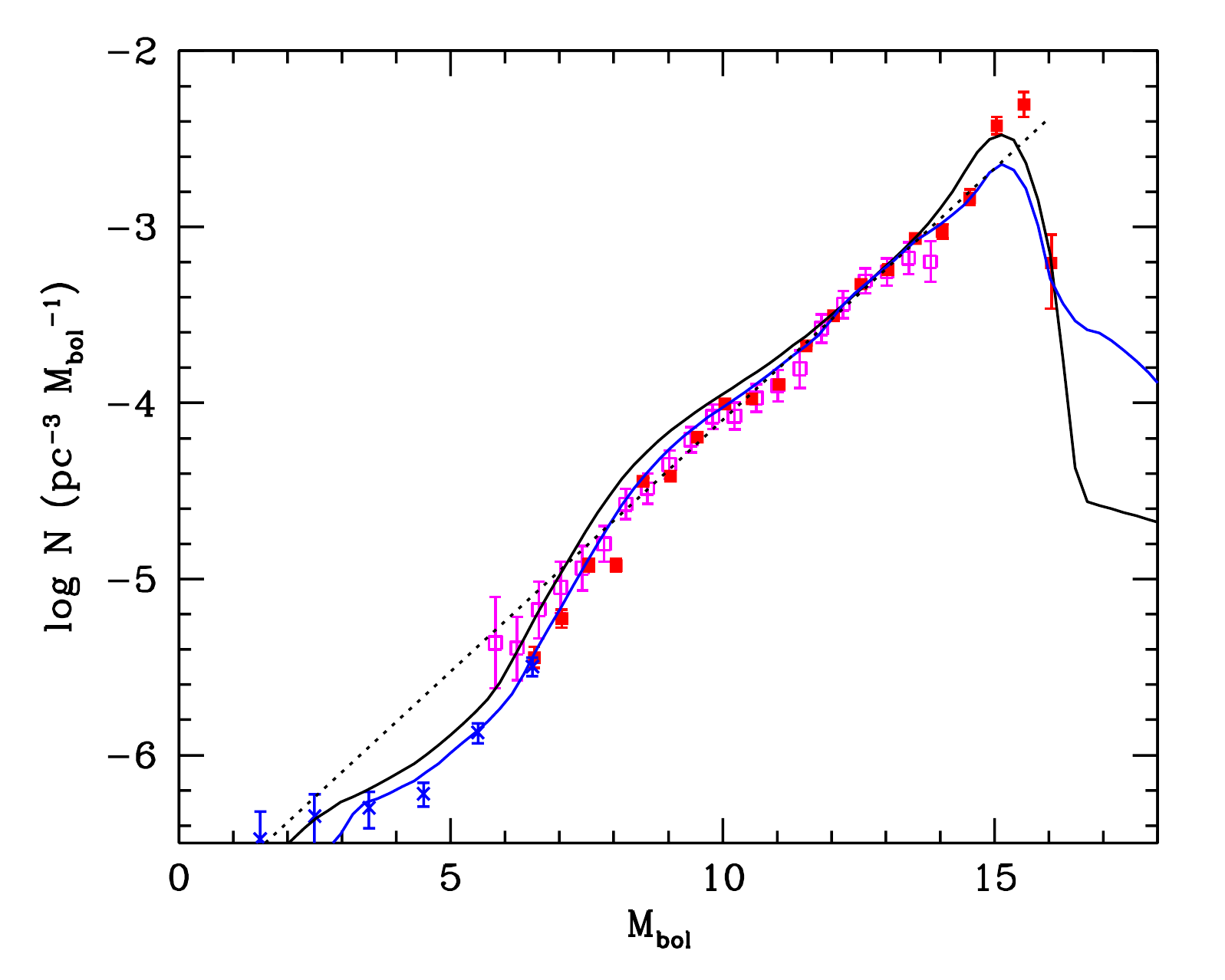}
\caption{\textit{Luminosity function of white dwarfs.}  Red (Harris \textit{et al.}~\cite{Harris:2005gd}) and blue (Krzesinski \textit{et al.}~\cite{Krzesinski:2009}) points represent the luminosity 
function of all white dwarfs (DA and non-DA families).  Magenta points~\cite{DeGennaro:2007yw} represent the luminosity function of the DA 
white dwarfs alone.  Both distributions have been normalized around $\Mbol= 13$, see text.  The dotted line represents the luminosity 
function obtained assuming Mestel's approximation.  The continuous lines correspond to full simulations assuming a constant star formation rate 
and an age of the Galaxy of 13 Gyr for the DA family (black line) and all, DA and non-DA, white dwarfs (blue line).}
\label{FigLF}
\end{figure}

It is important to realize that white dwarfs are divided into two broad categories, DA and non-DA. The DA white dwarfs exhibit hydrogen lines in 
their spectra caused by the presence of an external layer made of almost pure H. This hydrogen layer is absent in the case of non-DAs and, 
consequently, their spectra is free of the H spectral features. The main result is that the DAs cool down more slowly than the non-DAs~\cite{Althaus:2010pi}.

The observed LF is shown in Fig.~\ref{FigLF} for three different datasets. Note that moving from left to right along the horizontal axis we go from 
high luminosity (hot, young WDs) to low luminosity (cold, old WDs). The Harris \textit{et al.}~\cite{Harris:2005gd} (red) and the Krzesinski \textit{et al.}~\cite{Krzesinski:2009} (blue) data are representative of all, DAs and non-DAS, white dwarfs. The Harris \textit{et al.} LF has been constructed using the 
reduced proper motion method which is accurate for cold WDs with $\Mbol\gtrsim6$ but not appropriate for hot WDs with $\Mbol\lesssim6$, and which 
have been thus removed from the sample. The Krzesinski \textit{et al.} LF on the other hand has been built employing the UV-excess technique which 
is accurate for hot WDs with $\Mbol\lesssim7$ but inappropriate for the colder ones. Since the datasets overlap and, assuming continuity, it is possible 
to construct a LF that extends from $\Mbol\sim1.5$ to $\Mbol \sim16$, although the cool end is affected by severe 
selection effects. The DeGennaro \textit{et al.}~\cite{DeGennaro:2007yw} sample was also obtained with the proper motion technique, which is why the hot end is not reliable and has been removed. Since the identification of DAs and non-DAs is not clear at low temperatures, the corresponding points of~\cite{DeGennaro:2007yw} have 
been removed from Figure~\ref {FigLF}.  See Isern \textit{et al.}~\cite{Isern:2012xf} for a detailed discussion. Since ultimately only the slope of the 
LF is of interest and the total density of WDs is quite uncertain, it is usually more convenient to normalize the LF with respect to one of its values, 
which is commonly chosen around $\log(L/\Lsun)=-3$.

If the cooling were due only to photons and one assumes that Mestel's approximation~\cite{1952MNRAS.112..583M} holds (\textit{i.e.} ions 
behave like an ideal gas and the opacity of the radiative envelope follows Kramer's law), then the LF would be a straight line on this logarithmic 
plot, which already provides a reasonable fit to the data.  Note, however, that the data show a dip for values of $\Mbol$ around $6-7$.  That is 
where the neutrinos enter the game: for the hotter WDs (to the left in Fig.~\ref{FigLF}), neutrino emission becomes more 
important than photon cooling. When neutrinos are included and the cooling is simulated with a full stellar evolution code the agreement becomes 
impressive (see the continuous lines of Fig.~\ref{FigLF}). This agreement can be used to bound the inclusion of new sources or sinks of energy~\cite{Isern:2008fs}.

%The description just given is only qualitative and is meant to highlight the features of the LF that will play an important role in constraining BSM 
%models.  For a quantitative analysis one needs to employ a full stellar evolution code and modify it to take into account new possible cooling 
%mechanisms.

%It is useful to understand what the dominant production mechanisms for light bosons and light fermions are inside WDs, and how their energy 
%loss rates depend differently on the internal temperature, in order to assess their relative impact on the LF.  This is the subject of our next section.

%%%%%%%%%%%%%%%%%%%%%%%%%%%%%%%%%%%%%%%%%%%%%%%%%%%%%%%%%%%%%%%%%%%%%%%
%%%%%%%%%%%%%%%%%%%%%%%%%%%%%%%%%%%%%%%%%%%%%%%%%%%%%%%%%%%%%%%%%%%%%%%
%%%%%%%%%%%%%%%%%%%%%%%%%%
%%%%%%%%%%%%%%%%%%%%%%%%%%%%%%%%%%%%%%%%%%%%%%%%%%%%%%%%%%%%%%%%%%%%%%%
%%%%%%%%%%%%%%%%%%%%%%%%%%%%%%%%%%%%%%%%%%%%%%%%%%%%%%%%%%%%%%%%%%%%%%%
%%%%%%%%%%%%%%%%%%%%%%%%%%

\section{Cooling mechanisms}\label{Cooling}

In this section we review the various cooling mechanisms for WDs.  The aim is to provide the reader with a simple understanding of what 
mechanism dominates in what regime.

\subsection{Photons}

\begin{figure}[!t]
\centering
\includegraphics[width=0.6\textwidth]{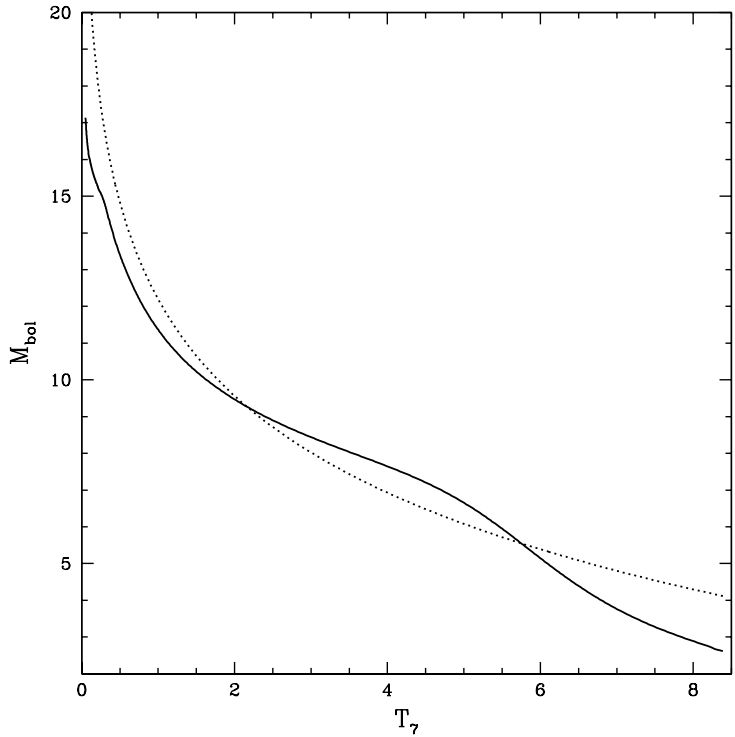}
\caption{Luminosity (bolometric magnitude) versus core temperature for a realistic model (continuous line) and for Mestel's model (dashed line). 
The photon luminosity $L_\gamma \simeq \epsilon_\gamma M_{\rm WD}$, with $M_{\rm WD}$ the WD mass, is related to $\Mbol$ via Eq.~(\ref{EqMbol}).} 
\label{FigLTc}
\end{figure}

In WDs the thermal energy is mostly stored in the nuclei which form, to a good approximation, a classical Boltzmann gas.  Taking into account the 
thermal conductance of the surface layers, one can relate the rate of energy loss at the surface to the internal temperature.  Using Mestel's 
approximation~\cite{1952MNRAS.112..583M} one finds
\eqn{
\epsilon_\gamma=3.29\times10^{-3}\ T_7^{7/2}\ \text{erg}\ \text{g}^{-1}\ \text{s}^{-1},
}[EqPhotons]
where $\epsilon_\gamma$ is the energy-loss rate per unit mass and $T_7\equiv\frac{T}{10^7\,\text{K}}$. This constitutes the main cooling for cold 
($\Mbol\gtrsim7$) WDs.  Realistic models indicate that $\epsilon_\gamma\propto T^{\beta}$, where $\beta \approx 7/2$, but varies slightly with temperature, chemical composition and mass of the white dwarf. We show in Fig.~\ref{FigLTc} what this energy loss as a function of the core temperature looks like for a realistic model as opposed to Mestel's model.

\subsection{Light bosons vs light fermions in white dwarfs}

Additional light bosons and light fermions that interact very weakly can also contribute to the cooling of WDs, but their dominant production mechanisms are 
usually different. 
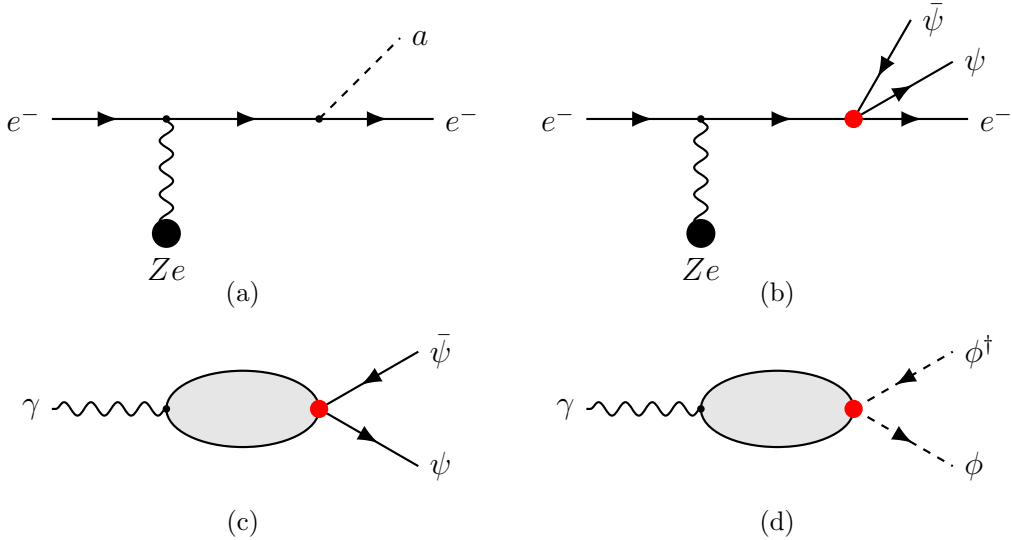
\begin{figure}[!t]
\centering
\resizebox{14cm}{!}{
\begin{tikzpicture}[thick]
\begin{scope}
\draw[-<-](1.5,0)--(0,0) node[left]{$e^-$};
\draw[->-](1.5,0)--(3.5,0);
\draw[->-](3.5,0)--(5,0) node[right]{$e^-$};
\draw[solid,decorate,decoration=snake](1.5,0)--(1.5,-1.5);
\filldraw[black](1.5,-1.5) circle(5pt) node[below,yshift=-5pt]{$Ze$};
\draw[dashed](3.5,0)--+(45:1.5cm) node[right]{$a$};
\filldraw[black](1.5,0) circle(1pt);
\filldraw[black](3.5,0) circle(1pt);
\draw (2.5,-2.3) node{\footnotesize{(a)}};
\end{scope}
\begin{scope}[xshift=7cm]
\draw[-<-](1.5,0)--(0,0) node[left]{$e^-$};
\draw[->-](1.5,0)--(3.5,0);
\draw[->-](3.5,0)--(5,0) node[right]{$e^-$};
\draw[solid,decorate,decoration=snake](1.5,0)--(1.5,-1.5);
\filldraw[black](1.5,-1.5) circle(5pt) node[below,yshift=-5pt]{$Ze$};
\draw[-<-](3.5,0)--+(60:1.5cm) node[right]{$\bar{\psi}$};
\draw[->-](3.5,0)--+(30:1.5cm) node[right]{$\psi$};
\filldraw[black](1.5,0) circle(1pt);
\filldraw[red](3.5,0) circle(3pt);
\draw (2.5,-2.3) node{\footnotesize{(b)}};
\end{scope}
\begin{scope}[yshift=-3.8cm]
\draw[solid,decorate,decoration=snake](1.5,0)--(0,0) node[left]{$\gamma$};
\filldraw[gray!20!white](2.5,0) ellipse(1cm and 0.5cm);
\draw[black](2.5,0) ellipse(1cm and 0.5cm);
\draw[-<-](3.5,0)--+(30:1.5cm) node[right]{$\bar{\psi}$};
\draw[->-](3.5,0)--+(330:1.5cm) node[right]{$\psi$};
\filldraw[black](1.5,0) circle(1pt);
\filldraw[red](3.5,0) circle(3pt);
\draw (2.5,-1.5) node{\footnotesize{(c)}};
\end{scope}
\begin{scope}[xshift=7cm,yshift=-3.8cm]
\draw[solid,decorate,decoration=snake](1.5,0)--(0,0) node[left]{$\gamma$};
\filldraw[gray!20!white](2.5,0) ellipse(1cm and 0.5cm);
\draw[black](2.5,0) ellipse(1cm and 0.5cm);
\draw[dashed,-<-](3.5,0)--+(30:1.5cm) node[right]{$\phi^\dagger$};
\draw[dashed,->-](3.5,0)--+(330:1.5cm) node[right]{$\phi$};
\filldraw[black](1.5,0) circle(1pt);
\filldraw[red](3.5,0) circle(3pt);
\draw (2.5,-1.5) node{\footnotesize{(d)}};
\end{scope}
\end{tikzpicture}
}
\caption{\textit{Processes for the production of new light scalars and fermions in WDs.}  The upper two diagrams represent bremsstrahlung 
processes, while the lower two diagrams show the plasmon decay.  The latter is the main production mechanism in WDs when the new light 
particles are produced in pairs.  The red dot in these diagrams represents an effective interaction given by a dimension 6 operator.  The grey blob 
in the lower two diagrams represents the effects of the stellar medium, as discussed in the text.}
\label{FigDiagrams}
\end{figure}
First, consider the DFSZ axion~\cite{Dine:1981rt,Zhitnitsky:1980tq} as an example of a light boson.  It would be mainly produced by the 
bremsstrahlung process $e+(Z,A)\to e+(Z,A)+a$ as 
shown in Fig.~\ref{FigDiagrams}~(a).  Raffelt gave an intuitive argument~\cite{Raffelt:1985nj} to understand how the corresponding energy 
emission rate depends on the temperature.  It goes as follows: The relevant interaction term in the Lagrangian is $iga\bar{e}\gamma_5e$, where $g=m_e/f_
\text{PQ}$, with $m_e$ the electron mass, $f_\text{PQ}\ge 10^9$ GeV the Peccei-Quinn scale, $a$ the axion field and $e$ the electron field.  The 
axion 
emission by an electron is analogous to the emission of a photon but, due to the presence of the $\gamma_5$, there is an extra electron spin-flip in the 
amplitude.  
Whereas the usual photon bremsstrahlung cross section is proportional to $E_\gamma^{-1}$, the axionic analogue is proportional to $E_a$ due 
to 
the extra power $E_a^2$ from the spin-flip nature of the process.  For the energy emission rate, we have to multiply the cross section by another 
factor of $E_a$, which makes it proportional to $E_a^2$.  We still have to do the phase space integrals for the initial and final state electrons. 
Because electrons are degenerate in WDs, these integrals contribute a factor of $T/E_F$ each, with $E_F$ the electron Fermi energy.  Combining the factors, 
the emission rate is proportional to $E_a^2(T/E_F)^2\propto T^4$, given that axion energies will be of the order of the temperature $T$.

Next, consider the bremsstrahlung of two fermions $\psi$, as depicted in Fig.~\ref{FigDiagrams}~(b).  The intuitive reasoning is 
analogous to what we just described for axions, but the difference is that the fermions from the electron line are produced in pairs (angular momentum conservation) as opposed to the single axion.  This adds an extra factor of the energy ($\sim T$) in the cross section and an extra phase space integral.  As a result, we have 
two more powers of $T$ in the final emission rate, which is therefore proportional to $T^6$.

When the calculations are done carefully one gets the following results for the energy-loss rates per unit mass in the two cases~\cite{Raffelt:1996wa}
\eqna{
\epsilon_a^\text{brem} &=1.08\times10^{-3}\ \alpha_{26}\ T_7^4\ \sum_j X_j\frac{Z_j^2}{A_j}\ F_a\ \text{erg}\ \text{g}^{-1}\ \text{s}^{-1},\\
\epsilon_\psi^\text{brem} &=1.34\times10^{-7}\left(\frac{C_\psi G_\psi}{C_VG_\text{F}}\right)^2T_7^6\ \sum_j X_j\frac{Z_j^2}{A_j}\ F_\psi\ \text{erg}\ 
\text{g}^{-1}\ \text{s}^{-1}.
}[EqBrem]
Here, $\alpha_{26}\equiv10^{26}\frac{g^2}{4\pi}$, with $g$ the coupling of the axion to electrons defined above;\footnote{For $\alpha_{26}$ of 
order one, axion cooling becomes comparable to photon cooling and one gets a better fit to the LF~\cite{Isern:2012ef}.  This fact can be taken as 
tentative evidence for the existence of axions, and it explains the choice of the power of 26 in the definition of $\alpha_{26}$.} $G_\psi$ is the 
dimensionful coupling for the four-fermion interaction denoted by a red dot in Fig.~\ref{FigDiagrams}, to be compared to the familiar Fermi 
constant, $G_\text{F}=1.166\times10^{-5}$ GeV$^{-2}$; $C_\psi$ is the effective coupling constant analogous to the effective neutral-current 
vector 
coupling constant $C_V=0.964$; $X_j$ is the mass fraction of the element $j$, with nuclear charge $Z_j$ and atomic mass number $A_j$, and 
the 
sum runs over the species of nuclei present in the WD.  $F_a$ and $F_\psi$ are factors that take into account the effect of screening for Coulomb 
scattering in a plasma.  In WDs $F_a$ and $F_\psi$ are of order one to a good approximation.

\begin{figure}[!t]
\centering
%\resizebox{6cm}{!}{
\begin{tikzpicture}
\begin{semilogyaxis}[xlabel=$T_7$,ylabel={$\epsilon$ [erg g$^{-1}$ s
$^{-1}$]},xmin=1,xmax=15,ymin=0.01,ymax=1000,width=10cm,height=7.5cm,legend style={legend pos=outer north east}]
\addplot[blue,solid,thick,domain=1:15]{3.29E-3*x^(3.5)};
\addlegendentry{Photon}
\addplot[red,solid,thick,domain=1:15]{1.40E+15*(1)*(1.69E-3*x)^9*(28*x^(-1))^6*exp(-(28*x^(-1)))*(2.4+0.6*(28*x^(-1))^(0.5)+0.51*(28*x^(-1))
+1.25*(28*x^(-1))^(1.5)+(8.6*(28*x^(-1))^2+1.35*(28*x^(-1))^(3.5))*(225-7*(28*x^(-1))+(28*x^(-1))^2)^(-1))};
\addlegendentry{Plasmon ($\bar{\psi}\psi$)}
\addplot[green,dashed,thick,domain=1:15]{1.08E-3*0.3*(1)*x^4*3.5};
\addlegendentry{Bremsstrahlung ($a$)}
\addplot[orange,dashed,thick,domain=1:15]{1.34E-7*(1)*x^6*3.5};
\addlegendentry{Bremsstrahlung ($\bar{\psi}\psi$)}
\end{semilogyaxis}
\end{tikzpicture}
%}
\caption{\textit{Comparison of energy losses in WDs.}  In this plot, $T_7 \equiv \frac{T}{10^7 \ {\rm K}}$, we have set $\alpha_{26}=0.3$, $C_\psi 
G_
\psi=C_VG_\text{F}$, and we have assumed a WD composed of an equal mixture of $^{12}$C and $^{16}$O.  For two fermions in the final state, 
the 
plasmon contribution dominates at high internal temperature, $T_7>5$, while the bremsstrahlung contribution is completely negligible.}
\label{FigCompareloss}
\end{figure}
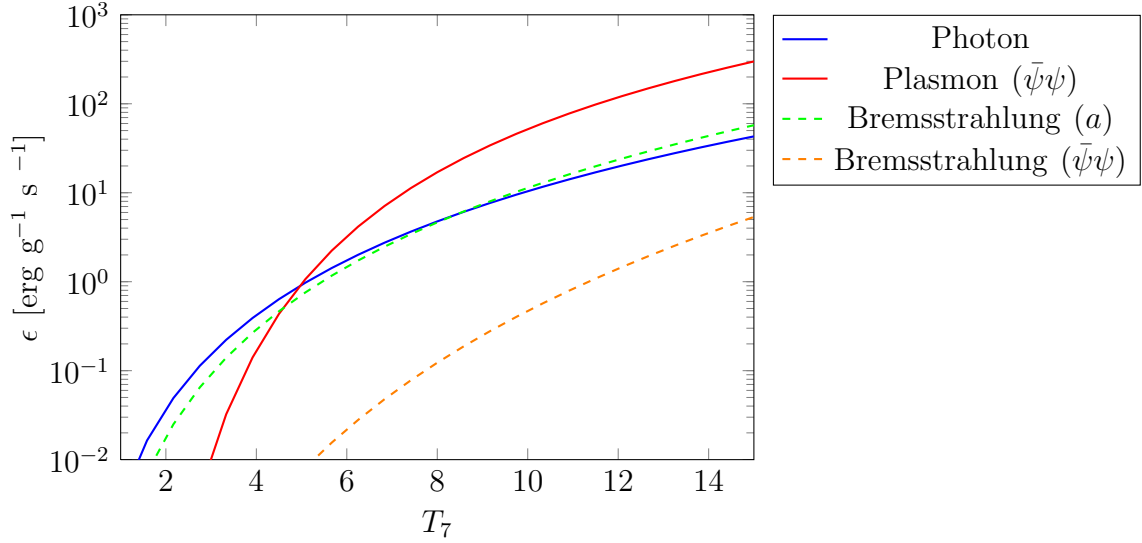

It is clear from expressions (\ref{EqPhotons}) and (\ref{EqBrem}) why the bremsstrahlung process for the neutrinos, where $C_\psi G_\psi=C_V 
G_
\text{F}$, is completely irrelevant in WDs, with internal temperature of the order of $10^7$ K.  First, the numerical coefficient in $\epsilon_\psi^\text{brem}$ is 
suppressed by four orders of magnitude compared to photons (and to axions if we take $\alpha_{26}$ of order one).  Second, it has a steeper 
dependence on the temperature, which makes it less and less relevant as we go to lower temperatures (see Fig.~\ref{FigCompareloss}).
Unless we have a model in which $G_\psi$ is significantly bigger than $G_\text{F}$, this contribution is negligible.  In fact, the dominant 
production 
mechanism of a pair of light fermions in WDs is not bremsstrahlung but is given by the so-called plasmon process~\cite{1963PhRv..129.1383A}, 
which is depicted in Fig.~\ref{FigDiagrams}~(c).  That is what we describe next.

In vacuum the photon is massless and can not decay into a pair of massive particles, no matter how light they are.  But in a medium, as in the 
interior 
of a star, the photon dispersion relations are modified and this allows such a decay.  What happens is that the photon also acquires a longitudinal 
polarization and is promoted 
to the so-called plasmon.  One would be tempted to say that the photon becomes massive, but such a statement is strictly speaking incorrect.  A 
better way to think about the plasmon decay, without ever referring to the mass of the photon, is the following: the propagation of an 
electromagnetic excitation (the plasmon) in the plasma is accompanied by an organized oscillation of the electrons, which in turn serve as a source 
for emitting a pair of light particles.  Figs.~\ref{FigDiagrams}~(c,d) are then understood as follows: the grey blob represents 
the medium response to the electromagnetic excitation; we can think of the black line outlining the blob as a loop of electrons, with the red dot 
denoting an effective interaction with the pair of light particles, that can be either fermions or bosons.  This is a schematic description.  The reader 
interested in more details is referred to the pedagogic treatment in chapter 6 of Ref.~\cite{Raffelt:1996wa}.

The calculation of the plasmon decay~\cite{1963PhRv..129.1383A, 1965NCimA..40..502Z} is quite involved, due to the effects of the medium, 
and 
cannot be performed analytically.  However, a good approximation, in the case of neutrinos as the products of the decay, was given in Ref.~\cite{Haft:
1993jt}.  The result applies to a wide range of stellar temperatures and densities.  Restricting ourselves to WDs, we can write it as
\eqn{
\epsilon_\psi^\text{plasmon}=1.40\times10^{15}\left(\frac{C_\psi G_\psi}{C_VG_\text{F}}\right)^2\ \lambda^9\gamma^6e^{-\gamma}(f_T+f_L)\ 
\text{erg}\ \text{g}^{-1}\ \text{s}^{-1},
}[EqEpsplasmon]
where numerically, to a good approximation
\eqn{
\lambda=1.69\times10^{-3}\ T_7,\quad\quad\gamma=\frac{28}{T_7},
}[Eqlambdagamma]
and
\eqn{
f_T=2.4+0.6\gamma^{1/2}+0.51\gamma+1.25\gamma^{3/2},\quad\quad f_L=\frac{8.6\gamma^2+1.35\gamma^{7/2}}{225-17\gamma+\gamma^2}.
}[Eqfs]

The plasmon decay depends in a complicated way on the photon dispersion relation in the medium.  However its main features can be understood in 
an approximation where the photons are treated as particles with an effective mass equal to the plasma frequency, $\omega_p$, which in the 
zero-temperature limit is given by~\cite{Raffelt:1996wa} $\omega_p^2 = 4\pi \alpha n_e/E_F$, with $\alpha$ the fine-structure constant, $n_e$ 
the electron density and $E_F$ the Fermi energy of the electrons. $\omega_p$ is of the order of a few tens of keV in WDs, slightly higher than the 
typical WD internal temperature, which is a few keV~\cite{Raffelt:1996wa}.  For the plasmon decay to happen, the decay products have to be 
kinematically accessible.  Thus, when we talk about {\em new light particles} in this context we mean particles {\em lighter than a few tens of keV}.

It is not immediately obvious how the energy loss of Eq.~\EqEpsplasmon compares to the previous ones because of its complicated form, but the 
differences can be easily visualized in the simple plot in Fig.~\ref{FigCompareloss}. For WDs whose internal temperature is below $4-5\times 
10^7$ K, the cooling is dominated by photons, and perhaps axions. Above that temperature, the plasmon decay into two light particles becomes 
the main source of energy loss. The contribution from the bremsstrahlung of a pair of fermions is always negligible on the plot.

It is useful to translate from temperature to $\Mbol$. From Eq.~\EqPhotons, multiplying by a typical WD mass, $M_{\rm WD}$, that we take to be 0.6 
solar masses, we obtain the photon luminosity, $L_\gamma = M_{\rm WD} \epsilon_\gamma$. Plugging it into Eq.~(\ref{EqMbol}) we obtain an 
expression that relates the temperature $T_7$ to $\Mbol$. Thus, a temperature of $4-5\times 10^7$ K corresponds to values of $\Mbol$ between 
6 and 7, which is indeed where we see the neutrino dip in Fig.~\ref{FigLF}. 
The plasmon decay into a pair of light particles constitutes the dominant cooling mechanism for $\Mbol<6-7$, the exact figures depending on 
the mass of the WD and the properties of the envelope.

Note that in models where a light boson couples to the electrons through a Yukawa coupling, the important cooling mechanism is the 
bremsstrahlung where the boson is produced singly, as in Fig.~\ref{FigDiagrams}~(a).  Such is the case for the DFSZ axion. In other models, 
instead, the light bosons couple indirectly to the electrons through a mediator and can only be produced in pairs, as for example when the 
bosons are charged under a new symmetry. This is the case for models with a dark sector, for instance, which we study in section~\ref{Examples}. 
The dominant production for these bosons is then no longer the bremsstrahlung, but the plasmon decay.

%%%%%%%%%%%%%%%%%%%%%%%%%%%%%%%%%%%%%%%%%%%%%%%%%%%%%%%%%%%%%%%%%%%%%%%
%%%%%%%%%%%%%%%%%%%%%%%%%%%%%%%%%%%%%%%%%%%%%%%%%%%%%%%%%%%%%%%%%%%%%%%
%%%%%%%%%%%%%%%%%%%%%%%%%%
%%%%%%%%%%%%%%%%%%%%%%%%%%%%%%%%%%%%%%%%%%%%%%%%%%%%%%%%%%%%%%%%%%%%%%%
%%%%%%%%%%%%%%%%%%%%%%%%%%%%%%%%%%%%%%%%%%%%%%%%%%%%%%%%%%%%%%%%%%%%%%%
%%%%%%%%%%%%%%%%%%%%%%%%%%

\section{A generic constraint on models with new light particles}\label{Generic}

This section describes a generic constraint on models with new light particles obtained from WD cooling and trapping.  We also discuss analogous constraints from red giants (RGs) and big bang nucleosynthesis (BBN).

\subsection{White dwarf cooling constraint}

In this section we discuss generic constraints from WD cooling due to plasmon decay into new light particles, that can be either fermions or 
bosons.  The only requirement is that they should be lighter than a few tens of keV, for the decay to be kinematically possible. As mentioned 
at the end of the previous section, such a process affects the LF for values of $\Mbol$ below $6-7$. Particle physics models in which new plasmon 
decay channels are open will potentially be in tension with the data, given the remarkable agreement between standard cooling mechanisms, 
that include neutrino emission, and the observed LF~\cite{Isern:2012xf}. We want to quantify how much the plasmon decay rate can deviate from 
the standard one, considering the neutrinos as the only decay products.

To achieve this goal, it is useful to introduce a unified formalism reminiscent of the Fermi interactions for fermions.  In order to compare with 
the standard plasmon decay into neutrinos, it is necessary to describe the relevant interaction between neutrinos $\nu$ and electrons $e$.  The 
interaction is given by
\eqn{
\mathscr{L}_\nu=-\frac{C_VG_\text{F}}{\sqrt{2}}[\bar{\nu}\gamma^\mu(1-\gamma_5)\nu](\bar{e}\gamma_\mu e),
}[EqLneutrino]
where the contribution from the effective neutral-current axial coupling constant $C_A$ is negligible for our purpose and can be ignored~\cite{Braaten:1993jw}. 
From this Lagrangian one can compute the plasmon decay rate into two neutrinos. The result is~\cite{1963PhRv..129.1383A,1965NCimA..
40..502Z,1972PhRvD...6..941D}
\eqn{
\Gamma_{\nu,s}=\frac{C_V^2G_\text{F}^2}{48\pi^2\alpha}\frac{Z_s\pi_s^3}{\omega_s},
}[EqPlasmneutrino]
where $\alpha$ is the fine-structure constant, $Z_s$ is the plasmon wavefunction renormalization, $\pi_s$ is the effective plasmon mass which 
enters in the dispersion relation $\omega^2-k^2=\pi_s(\omega,k)$ for a plasmon with frequency $\omega$ and wave vector $k$, and the 
subscript $s=\{T,L\}$ denotes the plasmon polarizations (transverse and longitudinal, respectively). The explicit forms for $\pi_T$ and $\pi_L$ are 
involved. 
They can be found, for example, in Ref.~\cite{Raffelt:1996wa}. We just point out for this discussion that $\pi_s$ is proportional to $\alpha$, 
so that $\Gamma_{\nu,s}$ goes to zero if we turn off the electromagnetic interaction, as expected. With the standard plasmon decay rate into 
neutrinos, Eq.~\EqPlasmneutrino, the energy-loss rate per unit mass is given by Eq.~\EqEpsplasmon with $C_\psi G_\psi=C_VG_\text{F}$, 
\textit{i.e.}
\eqn{
\epsilon_\nu^\text{plasmon}=\left.\epsilon_\psi^\text{plasmon}\right|_{C_\psi G_\psi=C_VG_\text{F}},
}[Eqplasmons]
and the contribution to the luminosity that appears in Eq.~(\ref{lwd}) is simply $L_\nu=M_\text{WD}\,\epsilon_\nu^\text{plasmon}$.

Let us now turn to new neutrino-like cooling mechanisms for WDs.  For BSM models with new light fermions $\psi$, the relevant interactions are  
given by
\eqn{
\mathscr{L}_\psi=-C_\psi G_\psi(\bar{\psi}\gamma^\mu\psi)(\bar{e}\gamma_\mu e),\hspace{1cm}\text{or}\hspace{1cm}\mathscr{L}_\psi=-C_\psi G_
\psi(\bar{\psi}\gamma^\mu\gamma^5\psi)(\bar{e}\gamma_\mu e)
}[EqLfermions]
which are the appropriate analogs of the four-fermion interaction.  The quantities $C_\psi$ and $G_\psi$ have been described above.  For new 
light bosons $\phi$ which must be produced in pairs [see Fig.~\ref{FigDiagrams}~(d)], the interaction is
\eqn{
\mathscr{L}_\phi=-2C_\phi G_\phi(i\phi^\dagger\overleftrightarrow{\partial}^\mu\phi)(\bar{e}\gamma_\mu e),
}[EqLbosons]
with $\phi^\dagger\overleftrightarrow{\partial}^\mu\phi \equiv \phi^\dagger (\partial^\mu \phi) - (\partial^\mu \phi^\dagger)\phi $.
The corresponding quantities are the effective coupling constant $C_\phi$ and the dimensionful parameter $G_\phi$, which is the analog of 
the Fermi constant.  For both interactions the plasmon decay into two new light particles is
\eqn{
\Gamma_{x,s}=\frac{C_x^2G_x^2}{48\pi^2\alpha}\frac{Z_s\pi_s^3}{\omega_s},
}[EqPlasmnew]
where $\{C_x,G_x\}$ are given by $\{C_\psi,G_\psi\}$ for new light fermions or $\{C_\phi,G_\phi\}$ for new light bosons.  The extra plasmon decay 
channel will lead to an extra energy-loss rate per unit mass as in Eq.~\EqEpsplasmon with $C_\psi G_\psi=C_xG_x$, \textit{i.e.}
\eqn{
\epsilon_x^\text{plasmon}=\left.\epsilon_\psi^\text{plasmon}\right|_{C_\psi G_\psi=C_xG_x},
}[Eqplasmonbosons]
and an extra contribution to the total luminosity given by $L_x=M_\text{WD}\,\epsilon_x^\text{plasmon}$, as in Eq.~(\ref{lwd}).  In the following the 
relevant constants will be denoted simply by $\{C_x,G_x\}$ both for new light fermions and bosons.

As already mentioned, in order not to upset the excellent agreement between standard WD cooling mechanisms~\cite{Isern:2012xf}, \textit{i.e.} 
from photon emission and neutrino emission (relevant only for hotter WDs), and observational data, we postulate that plasmon decay into new 
light particles must not account for more than the plasmon decay into neutrinos.  In the  massless limit, both for new particles as well as neutrinos, this 
constraint can be stated simply as [see Eqs.~\EqPlasmneutrino and \EqPlasmnew\!\!]
\eqn{
C_xG_x\lesssim C_VG_\text{F}.
}[EqConstraint]
In other words, any new sufficiently light particles (\textit{i.e.} which are effectively massless in WDs), that can be produced through plasmon 
decay in WDs and can escape from WDs, generate extra cooling.  This extra cooling must be subdominant compared to standard plasmon decay 
into neutrinos. To validate this order-one constraint, it is now necessary to properly quantify the agreement between the standard cooling 
mechanisms and observational data.

\begin{figure}[!t]
\centering
\includegraphics[width=0.7\textwidth]{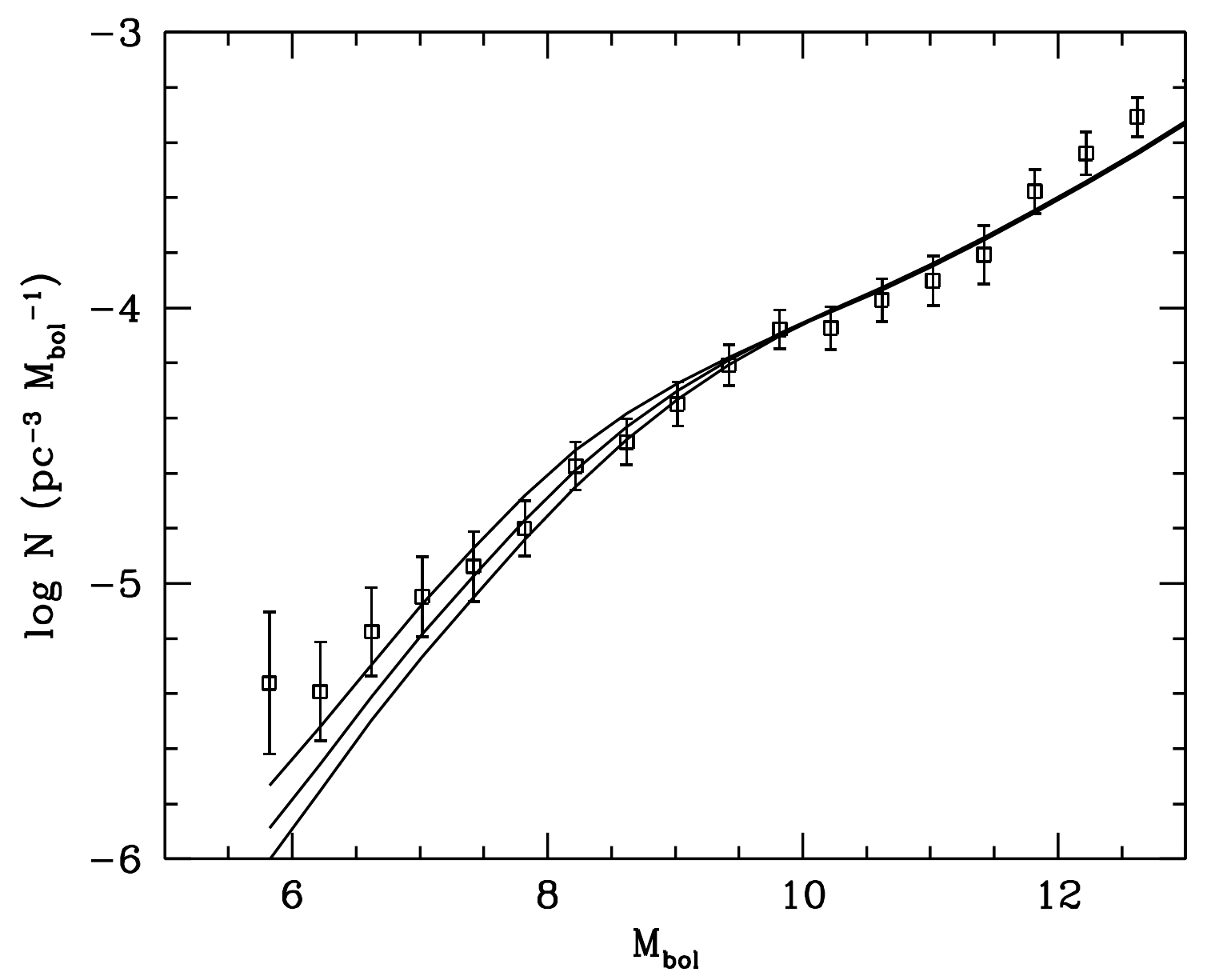}
\caption{\textit{Theoretical luminosity function for WDs.} The curves shown include the different contributions to Eq.~(\ref{lwd}) and correspond to values of $S_x=0,  0.5,  1$ from top to bottom for the $L_x = S_x L_\nu$ contribution. They are superimposed on the data points by DeGennaro \textit{et al.}~\cite{DeGennaro:2007yw}. }
\label{FigLFJordi}
\end{figure}

The standard cooling mechanisms relevant for WDs are photon cooling and plasmon decay into neutrinos.  Since we are interested in 
constraining models which lead to extra neutrino-like cooling, we focus here only on the dataset of DeGennaro \textit{et al.} which covers 
bolometric magnitudes between $5.5\lesssim\Mbol\lesssim12.5$.  This range is well understood and clearly exhibits the 
neutrino dip for $\Mbol$ around $6-7$ (see Fig.~\ref{FigLFJordi}).  Moreover, the dataset of DeGennaro \textit{et al.} has the smallest error bars in 
 this range and only contains DA WDs.

We start by minimizing the $\chi^2$ for the LF assuming standard cooling mechanisms and Mestel's approximation. The free parameter is the WD birthrate.
The best fit implies a birthrate $\sim 1.6\times 10^{-3}$ pc$^{-3}$ Gyr$^{-1}$, which is a reasonable local WD formation rate~\cite{Liebert:2004bv}, for 
$\chi_\text{min}^2=24.9$.  Since there are $N_\text{exp}=18$ data points and $N_\text{th}=1$ free parameters, this provides a decent fit with a
reduced chi-square $\chi_\text{red,min}^2=\chi_\text{min}^2/N_\text{dof}=1.47$, where $N_\text{dof}=N_\text{exp}-N_\text{th}=17$ is the total
 number of degrees of freedom.

Next we determine the 90\% confidence level exclusion contours for extra cooling from plasmon decay into new light particles, 
assuming the latter are massless.  Since the new plasmon decay channels are reminiscent of the standard plasmon decay into 
neutrinos, we take here $L_x=S_xL_\nu$, where $S_x$ determines the ratio of the new extra luminosity $L_x$ to the neutrino luminosity $L_\nu
 $. Now we take $S_x$ as our only free parameter, leaving the WD birthrate fixed to the value determined above. Thus we still have $N_\text{dof}=17$.
We then compute the new chi-square, $\chi^2$, including the $L_x$ contribution. From Fig.~36.1 in Ref.~\cite{Nakamura:2010zzi} we find that $\chi^2$ must be such that 
\eqn{
\Delta\chi^2=\chi^2-\chi_\text{min}^2<24.8
}[EqDeltachi]
otherwise the extra cooling is excluded at 90\% confidence level. Imposing the condition $\Delta\chi^2 <24.8$ translates into the constraint
\eqn{
S_x=\frac{L_x}{L_\nu}=\left(\frac{C_xG_x}{C_VG_\text{F}}\right)^2<0.99,
}[EqChiS]
which is equivalent to the one in Eq.~\EqConstraint, obtained from a simpler and more intuitive physical argument.

We provide in Fig.~\ref{FigLFJordi} three curves for the LF obtained from realistic models. The top one includes only standard cooling (with neutrinos), while the two lower ones include extra neutrino-like contributions with $S_x=0.5$ and $S_x=1$ respectively.

\subsection{White dwarf trapping constraint}

One should also include the effects of trapping.  Indeed, as $G_x$ increases the interactions between the new light particles and ordinary matter become stronger.  For very large $G_x$ the interactions are too strong and the mean free path of the new light particles is too small for them to escape the WD and thus contribute to its cooling. To make an estimate, we compare the cross section for the scattering of new light particles on ordinary matter, $\sigma_x \propto C_x^2 G_x^2$, with the corresponding one for neutrinos, $\sigma_\nu \propto C_V^2 G_\text{F}^2$.
Neutrinos have a mean free path of $\lambda_\nu=(n\sigma_\nu)^{-1}\simeq3000\Rsun$ in WDs~\cite{Althaus:2010pi}. Requiring that the mean free path of our light particles is bigger than a typical WD radius, $R_\text{WD}\simeq0.019\Rsun$~\cite{Raffelt:1996wa}, and comparing $\sigma_x$ and $\sigma_\nu$ we find the condition $C_xG_x\lesssim400 \ C_VG_\text{F}$. Combining this with Eq.~\EqConstraint implies that any new light particles produced in WDs are excluded by cooling considerations if
\eqn{C_VG_\text{F}\lesssim C_xG_x\lesssim 400 \ C_VG_\text{F}.
}[EqWDConstraint]
Eq.~\EqWDConstraint is the main result of this paper and will be used in section~\ref{Examples} to constrain BSM models with new light particles.

% Comparing with neutrinos, which have a mean free path of $\lambda_\nu=(n\sigma_\nu)^{-1}\simeq3000\Rsun$ in WD~\cite{Althaus:2010pi}, the trapping constraint for the new light particles can be written as $C_xG_x\gtrsim400C_VG_\text{F}$ assuming a WD with radius $R_\text{WD}\simeq0.019\Rsun$~\cite{Raffelt:1996wa}.  Combining constraint Eq.~\EqConstraint with the trapping constraint implies that any new light particles produced in WDs are excluded by cooling considerations if
%%
%\eqn{C_VG_\text{F}\lesssim C_xG_x\lesssim400C_VG_\text{F}.
%}[EqWDConstraint]
%%

\subsection{Comparison to constraints from red giants and big bang nucleosynthesis}

In the same line of thoughts, it is possible to obtain cooling constraints from red giants (RGs).  Following \cite{Raffelt:1996wa} the bound from RGs cooling can be translated into $S_x\lesssim2$, which corresponds to $C_xG_x\lesssim1.41C_VG_\text{F}$ and is comparable to, but slightly weaker than what we found in Eq.~\EqConstraint for WDs. 
 Moreover, since the cores of RGs can be seen as WDs, trapping constraints in RGs will necessary be worse than in WD. Therefore, in this context RGs do not constrain new light particles as well as WDs.

Such new light particles could however be very tightly constrained by BBN.  Given that we are interested in masses below a few tens of keV, if they were in thermal equilibrium with ordinary matter in the early universe until BBN, that happens at $T\sim 1$ MeV, they would contribute to the number of relativistic degrees of freedom, which is well constrained. To estimate this constraint, we follow~\cite{Steigman:2013yua}. 

The reactions $e^+ e^- \leftrightarrow \psi \psi$ and $e \psi \leftrightarrow e \psi$, responsible for keeping the light particle, $\psi$, in thermal equilibrium, have a typical cross section $\sigma_x \propto C_x^2G_x^2T^2$, which leads to an interaction rate per particle of $\Gamma_x=n\sigma_x|v|\propto C_x^2G_x^2T^5$, since their number density is $n\propto T^3$.  Comparing to the expansion rate $H\propto T^2/M_\text{Pl}$, the decoupling temperature can be estimated as $T_{x,\text{dec}}\propto(C_x^2G_x^2M_\text{Pl})^{-1/3}$, where $M_\text{Pl}$ is the Planck mass. This is completely analogous to the calculation for the neutrinos decoupling temperature, $T_{\nu,\text{dec}}\propto(C_V^2G_\text{F}^2M_\text{Pl})^{-1/3}$. Thus we can write
%
% The new light particle decoupling temperature $T_{x,\text{dec}}$ can be estimated by comparing to the neutrino decoupling temperature $T_{\nu,\text{dec}}$.  The typical cross section for such light particles is $\sigma_\nu\simeq C_V^2G_\text{F}^2T^2$ which leads to an interaction rate per particle of $\Gamma_\nu=n\sigma_\nu|v|\simeq C_V^2G_\text{F}^2T^5$ since $n\simeq T^3$ for relativistic particles.  Comparing to the expansion rate $H\simeq T^2/M_\text{Pl}$ the decoupling temperature can be estimated as $T_{\nu,\text{dec}}\simeq(C_V^2G_\text{F}^2M_\text{Pl})^{-1/3}$ where $M_\text{Pl}$ is the Planck mass.  The same can be done for the new light particles which leads to
%
\eqn{T_{x,\text{dec}}=\left(\frac{C_VG_\text{F}}{C_xG_x}\right)^{2/3}T_{\nu,\text{dec}}.
}[EqTdec]
Following~\cite{Steigman:2013yua} the effective number of neutrinos $N_\text{eff}$ is given by
\eqn{N_\text{eff}=3.018\left[1+\frac{\Delta N_\nu}{3}\left(\frac{10.73}{g_s(T_{x,\text{dec}})}\right)^{4/3}\right],
}[EqNeff]
where $g_s(T)$ is the ratio of the total entropy density to the photon entropy density and $\Delta N_\nu$ is the number of equivalent neutrinos, \textit{i.e.} $\Delta N_\nu=2\times1$ for a Dirac fermion or $\Delta N_\nu=2\times4/7$ for a complex scalar.  Demanding that the number of equivalent neutrinos be smaller than $4$ \cite{Ade:2013zuv} and taking $T_{\nu,\text{dec}}=3\ \text{MeV}$~\cite{Steigman:2013yua} leads to the constraint
\eqn{C_xG_x\lesssim(4.3\times10^{-3}\text{\,\,or\,\,}4.1\times10^{-2})C_VG_\text{F},
}[EqBBNbound]
which is three (two) orders of magnitude stronger than the WD bound Eq.~\EqConstraint for new light Dirac fermions (complex scalar bosons).  From this analysis it would thus seem that BBN bounds are more competitive than WD bounds in constraining models with new light particles.  Note, however, that there are caveats that could invalidate the BBN bounds without modifying the WD constraints.  For example, a light [$\sim\mathscr{O}(\text{MeV})$] weakly-interacting massive particle (WIMP) whose annihilations heat up the photons but not the neutrinos would result in a lower $N_\text{eff}$ and thus leave more room for extra relativistic degrees of freedom~\cite{Kolb:1986nf,Serpico:2004nm,Ho:2012br}.  In such a scenario, the bound of Eq.~\EqBBNbound would be relaxed to the extent that  the WD constraint would be more competitive.  Hence, the WD bound is robust because it is oblivious to possible caveats that would alter BBN considerations.

%%%%%%%%%%%%%%%%%%%%%%%%%%%%%%%%%%%%%%%%%%%%%%%%%%%%%%%%%%%%%%%%%%%%%%%
%%%%%%%%%%%%%%%%%%%%%%%%%%%%%%%%%%%%%%%%%%%%%%%%%%%%%%%%%%%%%%%%%%%%%%%
%%%%%%%%%%%%%%%%%%%%%%%%%%
%%%%%%%%%%%%%%%%%%%%%%%%%%%%%%%%%%%%%%%%%%%%%%%%%%%%%%%%%%%%%%%%%%%%%%%
%%%%%%%%%%%%%%%%%%%%%%%%%%%%%%%%%%%%%%%%%%%%%%%%%%%%%%%%%%%%%%%%%%%%%%%
%%%%%%%%%%%%%%%%%%%%%%%%%%

\section{Three examples}\label{Examples}

In this section we consider three examples of BSM scenarios.  The first two are supersymmetric extensions of the Standard Model (SM): in the 
first, the light particle is the neutralino, while in the second, it is the axino.  We show that WDs do not put competitive bounds on these 
models.  The situation is different in the third example, where we consider models with a dark sector, in which case the WDs bounds are very 
competitive.

\subsection{A light neutralino}

\begin{figure}[!t]
\centering
\resizebox{16cm}{!}{
\begin{tikzpicture}[thick]
\begin{scope}
\draw[-<-](2.5,2)--(0,2) node[left]{$e^-$};
\draw[->-](2.5,0)--(0,0) node[left]{$e^+$};
\draw[->-](2.5,2)--(5,2) node[right]{$\chi_0$, $\psi_a$};
\draw[-<-](2.5,0)--(5,0) node[right]{$\chi_0$, $\psi_a$};
\draw[dashed](2.5,0)--(2.5,2) node[midway,right]{$\tilde{e}$};
\filldraw[black](2.5,0) circle(1pt);
\filldraw[black](2.5,2) circle(1pt);
\draw (2.5,-0.75) node{\footnotesize{(a)}};
\end{scope}
\begin{scope}[xshift=7cm]
\draw[->](0,1)--(1,1);
\end{scope}
\begin{scope}[xshift=9cm]
\draw[-<-](2.5,1)--(0,2) node[left]{$e^-$};
\draw[->-](2.5,1)--(0,0) node[left]{$e^+$};
\draw[->-](2.5,1)--(5,2) node[right]{$\chi_0$, $\psi_a$};
\draw[-<-](2.5,1)--(5,0) node[right]{$\chi_0$, $\psi_a$};
\filldraw[red](2.5,1) circle(3pt) node[black,above]{$G_{\tilde{e}}$};
\draw (2.5,-0.75) node{\footnotesize{(b)}};
\end{scope}
\end{tikzpicture}
}
\caption{\textit{Processes for the production of light neutralinos or axinos in WDs.}  The first diagram represents the relevant production 
mechanism for plasmon decay into neutralinos or axinos through a selectron exchange.  The last diagram corresponds to the first diagram where 
the selectron is integrated out.  The red dot is the corresponding four-fermion effective interaction.}
\label{FigSUSY}
\end{figure}
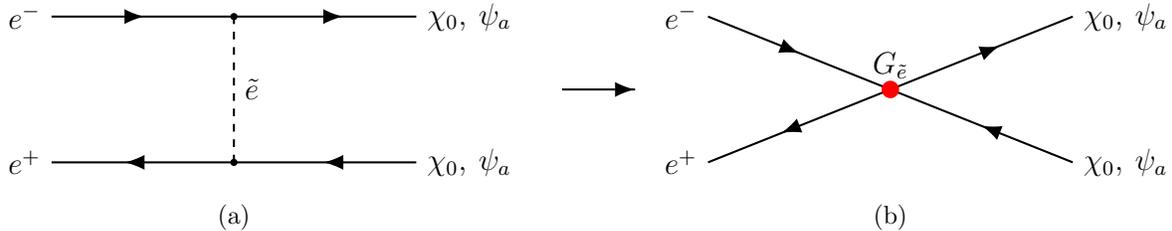

The neutralino $\chi_0$ is often the lightest supersymmetric particle in the Minimal Supersymmetric Standard Model.  It can be very light, even 
massless, and still evade all current experimental constraints~\cite{Dreiner:2009ic,Profumo:2008yg}.  For the production of light neutralinos in 
WDs, that would predominantly occur via plasmon decay, we consider the four-fermion interaction obtained from integrating out the selectron $
\tilde{e}$ (see Fig.~\ref{FigSUSY})\footnote{A very light neutralino, $m_{\chi_0}\ll 1$ GeV, is almost purely bino and does not couple to the $Z_0$~\cite{Choudhury}.},
\eqn{
\mathscr{L}_{\chi_0}=-C_{\chi_0}G_{\tilde{e}}(\bar{\chi}_0\gamma^\mu\gamma^5\chi_0)(\bar{e}\gamma_\mu e),
}[EqLneutralino]
where $G_{\tilde{e}}=\frac{e^2}{4\cos^2\theta_Wm_{\tilde{e}}^2}$ and $C_{\chi_0}=\frac{3}{4}$~\cite{Dreiner:2003wh}, with $e$ the electric 
charge, $\theta_W$ the weak mixing angle and $m_{\tilde{e}}$ the selectron mass.
Since $G_\text{F}=\frac{\sqrt{2}e^2}{8\sin^2\theta_Wm_W^2}$ where $m_W$ is the $W$ gauge boson mass the constraint Eq.~\EqConstraint can 
be translated into a lower bound on the selectron mass of
\eqn{
m_{\tilde{e}}\gtrsim\left(\frac{\sqrt{2}C_{\chi_0}}{C_V}\right)^{\frac{1}{2}}\tan\theta_Wm_W=45\ \text{GeV},
}[EqSelectron]
where $m_W=80.4$ GeV and $\sin^2\theta_W=0.23$.  Thus, in order to have a significant impact on the LF one needs a selectron lighter than the 
$W$ gauge boson.  The bound in Eq.~\EqSelectron applies to the case of a massless neutralino.  Turning on a small neutralino 
mass has the effect of pushing the WD bound down to even lower selectron masses.  Such light selectrons are already excluded by LEP
searches~\cite{Heister:2002jca}.  
Note that supernovae, contrary to WDs, provide a better arena to constrain the mass of a light neutralino~\cite{Dreiner:2003wh}.  Nevertheless, WD cooling 
bounds do not seem competitive for this process.

\subsection{A light axino}

A light axino is in principle very interesting in this context.  It has already been argued that the inclusion of an axion gives a better fit to the LF~\cite{Isern:2012ef}.  If 
supersymmetry (SUSY) is realized in nature, the axion would be necessarily accompanied by its fermionic partner, the axino, which could also be 
very light (see \textit{e.g.}~\cite{Rajagopal:1990yx, Chun, Chun199543}).  The axino could be pair-produced in the plasmon decay and contribute to the high luminosity part of 
the LF.  When combined with the contribution of the axion one might hope to get an even better fit.  Unfortunately, as we explain in the rest of this 
section, the axino interacts way too weakly so that its contribution to the LF turns out to be completely negligible.

Recall that the coupling of axions to electrons is given by $iga\bar{e}\gamma_5e$, with $g=m_e/f_\text{PQ}$.  In SUSY there is a corresponding 
axino-electron-selectron interaction that can be written as $ig\tilde{e}\bar{e}\psi_a$, where $\psi_a$ denotes the axino.  If we integrate out the 
selectron (see Fig.~\ref{FigSUSY}), the resulting four-fermion interaction between two electrons and two axinos is scalar-like [\textit{e.g.} $(\bar\psi_a 
\psi_a) (\bar e e)$]  instead of vector-like [\textit{e.g.} $(\bar\psi_a \gamma^\mu \psi_a) (\bar e \gamma_\mu e)$] 
and thus does not even allow plasmons to decay to pairs of axinos.  Being more precise and starting from the derivative interaction between the 
axion and electrons instead, one obtains higher-dimensional operators after supersymmetrizing and integrating out the selectron, \textit{i.e.} four-fermion interactions between two electrons and two axinos with extra derivatives, which are thus temperature-suppressed compared to the usual 
plasmon decay.  Most importantly however, these interactions are always at least suppressed by $g^2$, which is incredibly tiny for reasonable 
$f_\text{PQ}\sim10^9-10^{12}$ GeV.  Therefore, although the constraint Eq.~\EqConstraint cannot be applied directly here, the universal 
suppression just mentioned makes a possible production of axinos absolutely unobservable in WDs.

\subsection{A dark sector}
\subsubsection{The model}

As seen in the two previous examples, WD cooling might not seem to lead to any strong bounds on new light fermions.  The situation is however 
much more interesting when one considers models of BSM with massive dark photons~\cite{Fayet:1980ad,Fayet:1990wx}.  In these models, 
which 
could be of relevance as models of dark matter, a dark sector $\mathscr{L}_\text{D}$ communicates with the SM, $\mathscr{L}_\text{SM}$, solely 
through kinetic mixing $\mathscr{L}_{\text{SM}\otimes\text{D}}$~\cite{Holdom:1985ag}, \textit{i.e.}
\eqn{
\mathscr{L}=\mathscr{L}_\text{SM}+\mathscr{L}_\text{D}+\mathscr{L}_{\text{SM}\otimes\text{D}},\hspace{1cm}\text{where}\hspace{1cm}\mathscr{L}
_{\text{SM}\otimes\text{D}}=\frac{\varepsilon_Y}{2}F_{\mu\nu}^\text{SM}F_\text{D}^{\mu\nu}.
}[EqLdark]
Above the electroweak scale the kinetic mixing occurs with strength $\varepsilon_Y$ between the hypercharge gauge group $U(1)_Y$, with the 
corresponding $F_{\mu\nu}^\text{SM} = \partial_\mu B_\nu - \partial_\nu B_\mu$, and a new Abelian gauge group $U(1)_\text{D}$, with $F_\text{D}
^{\mu
\nu} = \partial^\mu A_\text{D}^\nu - \partial^\nu A_\text{D}^\mu$, where $A_\text{D}^\mu$ is the $U(1)_\text{D}$ gauge boson, \textit{i.e.} the dark photon.
Below the electroweak scale the mixing involves instead the electromagnetic gauge group, and $\varepsilon=\varepsilon_Y\cos\theta_W$.
The dimensionless parameter $\varepsilon$, which should be generated by integrating out massive states charged under both SM and dark 
gauge groups, is naturally small, $\varepsilon\sim10^{-4}-10^{-3}$.  Thus, after rotating the fields appropriately such that gauge bosons have 
canonically-normalized kinetic terms, the SM fields become millicharged under the {\em dark} gauge group~\cite{Cassel:2009pu,Hook:2010tw}, \textit{i.e.}
\eqn{
\mathscr{L}_{\text{SM}\otimes\text{D}}=-\varepsilon eJ_\mu^\text{SM}A_\text{D}^\mu,
}[EqLSMD]
where $J_\mu^\text{SM}$ is the SM electromagnetic current.

Thus, in models with massive dark photons, WD plasmons could decay, through off-shell massive dark photons, to light dark sector particles if 
they 
are kinematically available.  Note that both dark photon decays to bosons and fermions result in two-particle final states.  Thus such plasmon 
decay through massive dark photons into light dark sector particles is reminiscent of plasmon decay into fermions (\textit{e.g.}\ neutrinos) irrespective of 
the spin of the light dark sector particles [see Fig.~\ref{FigDiagrams}~(c,d)].  Therefore the relevant constraints for plasmon 
decay in models with massive dark photons are equivalent to the constraint discussed in section~\ref{Generic}.
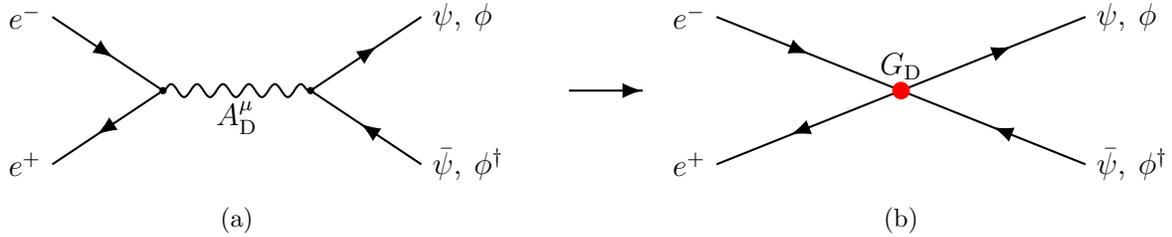
\begin{figure}[!t]
\centering
\resizebox{16cm}{!}{
\begin{tikzpicture}[thick]
\begin{scope}
\draw[-<-](1.5,1)--(0,2) node[left]{$e^-$};
\draw[->-](1.5,1)--(0,0) node[left]{$e^+$};
\draw[->-](3.5,1)--(5,2) node[right]{$\psi$, $\phi$};
\draw[-<-](3.5,1)--(5,0) node[right]{$\bar{\psi}$, $\phi^\dagger$};
\draw[solid,decorate,decoration=snake](1.5,1)--(3.5,1) node[midway,below]{$A_\text{D}^\mu$};
\filldraw[black](1.5,1) circle(1pt);
\filldraw[black](3.5,1) circle(1pt);
\draw (2.5,-0.75) node{\footnotesize{(a)}};
\end{scope}
\begin{scope}[xshift=7cm]
\draw[->](0,1)--(1,1);
\end{scope}
\begin{scope}[xshift=9cm]
\draw[-<-](2.5,1)--(0,2) node[left]{$e^-$};
\draw[->-](2.5,1)--(0,0) node[left]{$e^+$};
\draw[->-](2.5,1)--(5,2) node[right]{$\psi$, $\phi$};
\draw[-<-](2.5,1)--(5,0) node[right]{$\bar{\psi}$, $\phi^\dagger$};
\filldraw[red](2.5,1) circle(3pt) node[black,above]{$G_\text{D}$};
\draw (2.5,-0.75) node{\footnotesize{(b)}};
\end{scope}
\end{tikzpicture}
}
\caption{\textit{Process for the production of light dark sector particles in WDs.}  The first diagram represents the relevant production mechanism 
for 
plasmon decay into light dark sector particles through a dark photon exchange.  The last diagram corresponds to the first diagram where the dark 
photon is integrated out.  The red dot is the corresponding dimension 6 operator.}
\label{FigDark}
\end{figure}

We stress the fact that in the scenario we are contemplating, the dark $U(1)_{\rm D}$ gauge group is broken so that the dark photon is massive. Instead, when $U(1)_{\rm D}$ is unbroken, the corresponding gauge boson is commonly referred to as a paraphoton. In this latter case, dark sector particles acquire an electric millicharge, that is a tiny fractional charge under the visible $U(1)_{\rm EM}$, and the constraints are usually shown on the plane given by $\varepsilon$ versus the mass of the dark sector particle~\cite{Davidson:2000hf}. In our case, with the broken dark $U(1)_{\rm D}$, there are no particles with an electric fractional charge. Rather, SM particles have a fractional charge under $U(1)_{\rm D}$, that is quite different.

\subsubsection{Excluded parameter region}

In order to determine the resulting excluded parameter space it is necessary to integrate out the dark photon, as shown in Fig.~\ref{FigDark}.  
This 
leads to the interaction
\eqn{
\mathscr{L}_{\psi,\phi}=-G_\text{D}J_\text{D}^\mu J_\mu^\text{SM}\supset-G_\text{D}[C_\psi(\bar{\psi}\gamma^\mu\psi)+2C_\phi(i\phi^\dagger
\overleftrightarrow{\partial}^\mu\phi)](\bar{e}\gamma_\mu e),
}[Eqelectronsdark]
where the dark constant is $G_\text{D}=\frac{4\pi\varepsilon\sqrt{\alpha\alpha_\text{D}}}{m_{A_\text{D}}^2}$ and $C_\psi=Q_\psi$, $C_\phi=\frac{Q_
\phi}
{2}$.  Here $m_{A_\text{D}}$ is the dark photon mass, $\alpha_\text{D}$ is the dark fine-structure constant and $Q_{\psi,\phi}$ are the dark 
particle 
charges under the dark gauge group. Note that dark photon decay into a pair of dark gauge bosons is generically not kinematically accessible 
because the masses of the dark photon and of the other dark gauge bosons are usually of the same order, as, for example, the $Z$ and 
$W$ gauge bosons in the SM.

\begin{figure}[!t]
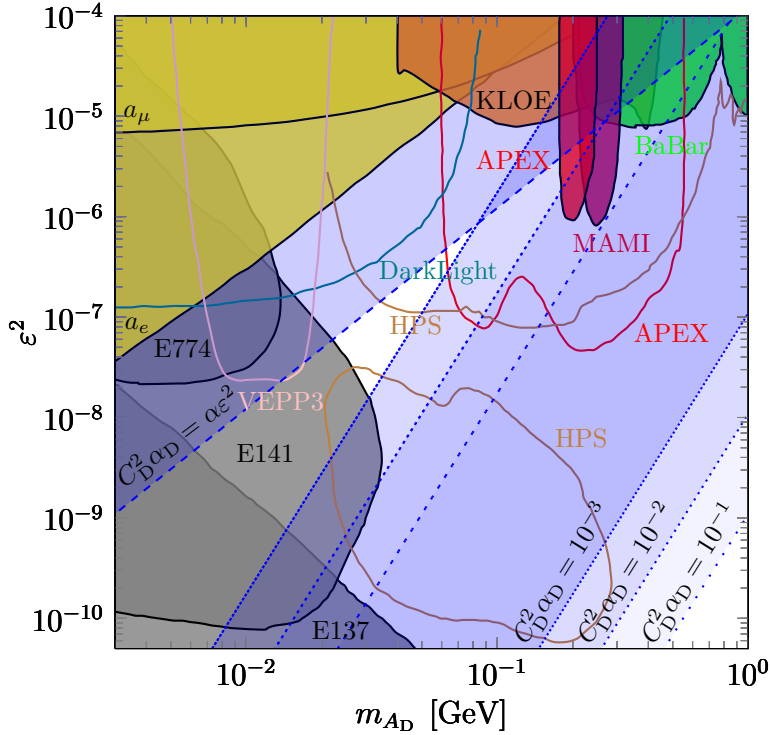

\centering
%\resizebox{6cm}{!}{

%}
\caption{\textit{Parameter space exclusion of dark forces with light ($\lesssim$ few tens of keV) hidden sector particles from energy losses in WDs.}  The blue shaded regions are excluded by WD cooling for $C_\text{D}^2\alpha_\text{D}=10^{-1}$ (loosely dotted lines), $C_\text{D}^2\alpha_\text{D}=10^{-2}$ (dotted lines) and $C_\text{D}^2\alpha_\text{D}=10^{-3}$ (densely dotted lines). On the left of these blue bands the hidden sector particles would be trapped inside the WD, which is why we cannot exclude that region with the simple cooling argument.  For experiments, which usually assume the dark photon decay is predominantly into the SM, shaded regions correspond to completed direct searches while curves show future reach.  For the electron and muon anomalous magnetic moments, shaded regions are excluded by measurements.  The reader is referred to the text for more details.}
\label{FigExclusion}
\end{figure}

Comparing with plasmon decay to neutrinos as discussed in section \ref{Generic}, the constraint Eq.~\EqWDConstraint leads to
\eqn{
1.09\times10^{-10}\left(\frac{m_{A_\text{D}}}{\text{GeV}}\right)^4=\frac{C_V^2G_\text{F}^2m_{A_\text{D}}^4}{16\pi^2\alpha}\lesssim C_\text{D}^2\alpha_\text{D}\varepsilon^2\lesssim\frac{400^2C_V^2G_\text{F}^2m_{A_\text{D}}^4}{16\pi^2\alpha}=1.09\times10^{-10}\left(\frac{20m_{A_\text{D}}}{\text{GeV}}\right)^4,
}[EqDark]
where $C_\text{D}=C_{\psi,\phi}$.  In Fig.~\ref{FigExclusion} we show the constraint Eq.~\EqDark and regions in parameter space which have 
already been explored or will be explored by future experiments~\cite{Bjorken:2009mm}, \textit{i.e.} beam dump experiments at SLAC: E137, 
E141 
and E774~\cite{Riordan:1987aw,Bross:1989mp,Andreas:2012mt}; $e^+e^-$ colliding experiments: BaBar~\cite{Aubert:2009au,Bjorken:
2009mm} 
and KLOE~\cite{Archilli:2011zc}; and fixed-target experiments: APEX~\cite{Abrahamyan:2011gv}, DarkLight~\cite{Freytsis:2009bh}, HPS~
\cite{Boyce:2012ym}, MAMI~\cite{Merkel:2011ze} and VEPP-3~\cite{Wojtsekhowski:2012zq}.  Fig.~\ref{FigExclusion} also shows excluded 
regions 
from electron ($a_e$) and muon ($a_\mu$) anomalous magnetic moment measurements~\cite{Pospelov:2008zw,Davoudiasl:2012ig,Endo:2012hp}.

For reasonable dark sector parameters where $\alpha_\text{D}\sim\alpha$, one has $C_\text{D}\sim1$ and $C_\text{D}^2\alpha_\text{D}\sim10^{-3}-10^{-2}$, and thus all but a small fraction of the relevant dark sector parameter space is excluded by WD cooling and the some of the above-mentioned experiments become obsolete \textit{if} dark photons couple to new dark sector fermions and/or bosons which are effectively massless in WDs, \textit{i.e.} lighter than a few keV.  In other words, to be viable models of dark photons, any model probed by the above-mentioned experiments cannot have dark sector fermions and/or bosons lighter than a few tens of keV due to WD cooling.

Note however that all of the experiments shown in Fig.~\ref{FigExclusion}---apart from DarkLight, VEPP-3 and the anomalous magnetic moment measurements---assume that dark photons decay predominantly back into the SM.  Although this is not possible in WDs (dark photons could only decay back into electron-positron pairs which are not kinematically accessible, the decay to neutrino pairs is negligible), this assumption forbids either light dark sector particles, in which case the WD constraint presented here is irrelevant; or large dark fine-structure constant (relative to $\alpha\varepsilon^2$), for which dark photon decay rate into invisible channel dominates.

To investigate this last possibility, we include in Fig.~\ref{FigExclusion} (see dashed blue line) the WD cooling constraint for which the dark photon decay rate into visible channels dominates over the decay rate into invisible channels, \textit{i.e.} $\Gamma_\text{invisible}\lesssim\Gamma_\text{visible}$ or $C_\text{D}^2\alpha_\text{D}\lesssim\alpha\varepsilon^2$.  It is interesting to see that, for very weak dark fine structure constant, the experiments which are sensitive to invisible dark photon decays, \textit{i.e.} DarkLight and VEPP-3, are still constrained by the WD cooling even when dark photons decay predominantly back into the SM.

Note that the constraint must be modified for a very light dark photon (again lighter than a few tens of keV), since it could be produced on-shell, which would result in an enhancement of the cooling rate. The resulting constraint would then be even tighter.  For such a light dark photon bremsstrahlung might also become important.

Finally, it would be of interest to study astrophysical cooling constraints from more energetic objects, like supernovae, to relax the restriction on the masses of the dark particles produced.

%%%%%%%%%%%%%%%%%%%%%%%%%%%%%%%%%%%%%%%%%%%%%%%%%%%%%%%%%%%%%%%%%%%%%%%
%%%%%%%%%%%%%%%%%%%%%%%%%%%%%%%%%%%%%%%%%%%%%%%%%%%%%%%%%%%%%%%%%%%%%%%
%%%%%%%%%%%%%%%%%%%%%%%%%
%%%%%%%%%%%%%%%%%%%%%%%%%%%%%%%%%%%%%%%%%%%%%%%%%%%%%%%%%%%%%%%%%%%%%%%
%%%%%%%%%%%%%%%%%%%%%%%%%%%%%%%%%%%%%%%%%%%%%%%%%%%%%%%%%%%%%%%%%%%%%%%
%%%%%%%%%%%%%%%%%%%%%%%%%

\section{Discussion and conclusion}

We studied constraints from the WD LF on BSM models with new light particles.  Whenever these light particles are produced in pairs, whether 
they are fermions or bosons, the dominant production mechanism in WDs is (usually) given by the plasmon decay.  Such a decay is responsible 
also for the production of neutrino pairs, whose effect is well understood and clearly visible through the dip at $\Mbol\sim6-7$ in the LF curve.  
Adding a significant decay into new light particles would deepen the dip, which would then be in disagreement with the data.  This constrains part 
of the parameter space of these BSM models.  More quantitatively, one needs to compare the strength of the interaction between the new light 
particles and the electrons with the interaction between neutrinos and electrons, \textit{i.e.} the Fermi constant $G_\text{F}$, and require that the 
former do not exceed the latter.

We applied this constraint to three models.  We first consider a supersymmetric model with a light neutralino and showed that the WD constraint is not competitive with existing collider bounds.  The situation is analogous with an axino, whose interaction is even further suppressed with respect to the neutralino, and does not lead to any interesting constraint.  We then explored models with a dark sector, for which the bounds are more relevant.  That is due mainly to the fact that the dark photon, that mediates the interaction between the electrons and the light dark sector particles, can be light [$\sim\mathscr{O}(\text{GeV})$], which enhances the plasmon decay rate.  It turns out that the limits on the dark sector parameter space from energy losses in WDs, as shown in Fig.~\ref{FigExclusion}, are extremely competitive and render some experiments obsolete \textit{if} the dark photon couples to light [$\sim\mathscr{O}(10\ \text{keV})$] dark sector particles.  Said differently, the dark photon models which are probed by these experiments cannot have light dark sector fermions and/or bosons, due to WD cooling.

Such dark sector particles could contribute to the relativistic degrees of freedom, $N_\text{eff}$, in the early universe and alter BBN predictions. BBN bounds can indeed be stronger than those from WD cooling. However, they are subject to caveats and are not as robust.

\vspace{5mm}

\noindent\textit{Note:} During completion of this work An \textit{et al.}~\cite{An:2013yfc} posted a paper on stellar constraints for dark photons.  
There is no overlap between our work and theirs since they consider dark photons with hard St\"{u}ckelberg masses.

%%%%%%%%%%%%%%%%%%%%%%%%%%%%%%%%%%%%%%%%%%%%%%%%%%%%%%%%%%%%%%%%%%%%%%%
%%%%%%%%%%%%%%%%%%%%%%%%%%%%%%%%%%%%%%%%%%%%%%%%%%%%%%%%%%%%%%%%%%%%%%%
%%%%%%%%%%%%%%%%%%%%%%%%%
%%%%%%%%%%%%%%%%%%%%%%%%%%%%%%%%%%%%%%%%%%%%%%%%%%%%%%%%%%%%%%%%%%%%%%%
%%%%%%%%%%%%%%%%%%%%%%%%%%%%%%%%%%%%%%%%%%%%%%%%%%%%%%%%%%%%%%%%%%%%%%%
%%%%%%%%%%%%%%%%%%%%%%%%%

\ack{We thank Rouven Essig for useful discussions and for comments on the manuscript. We also thank the referee for valuable comments.  HD and LU acknowledge the DFG SFB TR 33 ``The Dark Universe'' for support throughout this work. JFF is supported by the ERC grant BSMOXFORD No.\ 228169. JI is supported by the MINECO-FEDER grants AYA2011-24704/ESP, by the ESF EUROCORES Program EuroGENESIS (MINECO grant EUI2009-04170), and by the grant 2009SGR315 of the Generalitat de Catalunya.}

%%%%%%%%%%%%%%%%%%%%%%%%%%%%%%%%%%%%%%%%%%%%%%%%%%%%%%%%%%%%%%%%%%%%%%%
%%%%%%%%%%%%%%%%%%%%%%%%%%%%%%%%%%%%%%%%%%%%%%%%%%%%%%%%%%%%%%%%%%%%%%%
%%%%%%%%%%%%%%%%%%%%%%%%%
%%%%%%%%%%%%%%%%%%%%%%%%%%%%%%%%%%%%%%%%%%%%%%%%%%%%%%%%%%%%%%%%%%%%%%%
%%%%%%%%%%%%%%%%%%%%%%%%%%%%%%%%%%%%%%%%%%%%%%%%%%%%%%%%%%%%%%%%%%%%%%%
%%%%%%%%%%%%%%%%%%%%%%%%%

\bibliography{Neutralinos_Draft_Revised}

\end{document}